\documentclass[aps,prc,preprint,tightenlines,groupedaddress,nofootinbib,
showpacs,preprintnumbers,amsmath,amssymb,superscriptaddress]{revtex4}
\usepackage{graphicx}% Include figure files
\usepackage{dcolumn}%Align table columns on decimal point
\usepackage{bm}% bold math
\usepackage{amsmath}
\begin{document}
\title{Virtues and Flaws of the Pauli Potential}
\author{J. Taruna\footnote{\tt e-mail: yutri@fermi.physics.fsu.edu}
        J. Piekarewicz\footnote{\tt e-mail: jorgep@csit.fsu.edu}}
\affiliation{Department of Physics, Florida State University,
             Tallahassee, FL 32306}
\author{M.A. P\'{e}rez-Garc\'{\i }a\footnote{\tt e-mail: mperezga@usal.es}}
\affiliation{Departamento de F\'isica Fundamental and Instituto 
             Universitario de F\'isica Fundamental y Matematicas, 
             Plaza de la Merced s/n, Universidad de Salamanca E-37008 
             Salamanca, Spain}
\date{\today}

\vspace{0.5cm}

\begin{abstract}
Quantum simulations of complex fermionic systems suffer from a variety
of challenging problems. In an effort to circumvent these challenges,
simpler {\sl ``semi-classical''} approaches have been used to mimic
fermionic correlations through a fictitious {\sl ``Pauli
potential''}. In this contribution we examine two issues. First, we
address some of the inherent difficulties in a widely used version of
the Pauli potential. Second, we refine such a potential in a manner
consistent with the most basic properties of a cold Fermi gas, such 
as its momentum distribution and its two-body correlation function.
\end{abstract}
\maketitle 
%--------------------------------------------------------------------------
\section{Introduction}
\label{Sec:Intro}

Insights into the complex and fascinating dynamics of Coulomb
frustrated systems across a variety of disciplines are just starting
to emerge (see, for example, Ref.~\cite{Jamei:2005} and references
therein). In the particular case of neutron stars, one is interested
in describing the equation of state of neutron-rich matter across an
enormous density range using a single underlying theoretical model. In
recent simulations we have resorted to a classical model that, while
exceedingly simple, captures the essential physics of Coulomb
frustration and nuclear
saturation~\cite{Horowitz:2004yf,Horowitz:2004pv,Horowitz:2005zb}.
The model includes competing interactions consisting of a short-range
nuclear attraction (adjusted to reproduce nuclear saturation) plus a
long-range Coulomb repulsion. The charge-neutral system consists of
electrons, protons, and neutrons, with the electrons (which at these
densities are no longer bound) modeled as a degenerate free Fermi gas.

So far, the only quantum effect that has been incorporated into this
{\sl ``semi-classical''} model is the use of an effective temperature
to simulate quantum zero-point motion. The main justification behind
the classical character of the simulations is the heavy nature of the
nuclear clusters. Indeed, at the low densities of the neutron-star
crust, the de~Broglie wavelength of the heavy clusters is
significantly smaller than their average separation. However, this
behavior ceases to be true in the transition region from the inner
crust to the outer core. At the higher densities of the outer core,
the heavy clusters are expected to ``melt'' into a collection of
isolated nucleons with a de~Broglie wavelength that becomes comparable
to their average separation. Thus, fermionic correlations are expected
to become important in the crust-to-core transition
region. Unfortunately, in contrast to classical simulations that
routinely include thousands --- and even millions --- of particles,
full quantum-mechanical simulations of many-fermion systems suffer
from innumerable challenges (see Ref.~\cite{Grotendorst:2002} and
references therein). In an effort to ``circumvent'' --- although not
solve --- some of these formidable challenges, classical simulations
of heavy-ion collisions and of the neutron-star crust have resorted to
a fictitious {\sl ``Pauli potential''}.  Within the realm of nuclear
collisions, the first such simulations were those of Wilets and
collaborators~\cite{Wilets:1977}. Other simulations with a more
refined Pauli potential have
followed~\cite{Dorso:1987,Boal:1988,Peilert:1992,Maruyama:1997rp,Watanabe:2003},
but the spirit has remain the same: introduce a momentum dependent,
two-body Pauli potential that penalizes the system whenever two
identical nucleons get too close to each other in phase space.

A goal of the present contribution is to show that the demands imposed
by such a Pauli potential are too weak to reproduce some of the most 
basic properties of a zero-temperature (or cold) Fermi gas. Thus, we
aim at refining such a potential in a manner that three fundamental 
properties of a cold Fermi gas be reproduced. These are: ($i$) the 
kinetic energy (as others have done before us), ($ii$) the momentum 
distribution, and ($iii$) the two-nucleon correlation function. 

The manuscript has been organized as follows. In
Sec.~\ref{sec:formalism}, some of the basic properties of a free Fermi
gas are discussed. As our classical simulations must by necessity be
carried out at finite temperature, a Sommerfeld expansion is used to
compute low-temperature corrections to these observables. The section
concludes with a review of the {\sl ``standard''} form of the Pauli
potential, its flaws, and the measures that we take to overcome these
flaws. In Sec.~\ref{sec:results}, results for the kinetic energy,
momentum distribution, and two-body correlation function are presented
and contrasted against exact analytic results. Conclusions and
possible future directions are presented in
Sec.~\ref{sec:conclusions}.

%--------------------------------------------------------------------------
\section{Formalism}
\label{sec:formalism}

The present section starts with a brief overview of some fundamental
properties of a low-temperature Fermi gas~\cite{Pathria:1996}. Next, a
fictitious Pauli potential is introduced and constrained to reproduce
these properties via a purely classical simulation. In particular,
special emphasis is placed on the necessary modifications to the 
{\sl ``standard''} Pauli potential that are required to reproduce 
such fundamental properties.

\subsection{Free Fermi Gas}
\label{sec:FFG}

The zero temperature Fermi gas is the simplest many-fermion system.
Such a system displays no correlations beyond those imposed by the 
Pauli exclusion principle and is described by the following free 
Hamiltonian:
\begin{equation}
 H = \sum_{i=1}^{N} \frac{{\bf p}_{i}^{2}}{2m} \;.
 \label{FreeH}
\end{equation}
Here $m$ is the mass of the fermion and $N$ denotes the (large) number
of particles in the system. As no interaction of any sort exists among
the particles, the eigenstates of the system are given by a product of
(single-particle) momentum eigenstates, suitably antisymmetrized to
fulfill the constraints imposed by the Pauli principle. For
simplicity, we assume that the fermions reside in a very large box of
volume $V\!=\!L^{3}$ and that the momentum eigenstates satisfy
periodic boundary conditions. We will be interested in the thermodynamic
limit of $N\!\rightarrow\infty$ and $V\!\rightarrow\infty$, but with their 
ratio fixed at a specific value of the number density $\rho\!\equiv\!N/V$. 

The (``box'') normalized momentum eigenstates are simple plane waves.
That is,
\begin{equation}
  \varphi_{\bf p}({\bf r}) = 
  \frac{1}{\sqrt{V}}\,e^{i{\bf p}\cdot{\bf r}} \;. 
 \label{FreePhi}
\end{equation}
Given that the eigenvalue problem is solved in a finite box using periodic 
boundary conditions, the resulting single-particle momenta are quantized
as follows:
\begin{equation}
 {\bf p}({\bf n})=\frac{2\pi}{L}{\bf n}=
 \frac{2\pi}{L}(n_{x},n_{y},n_{z})\;,
 \;\;{\rm with}\;n_{i}=0,\pm1,\pm2,\ldots
 \label{PtoK}
\end{equation}
with the corresponding single-particle energies given by
$\epsilon({\bf p})= {\bf p}^{2}/2m$.

Up to this point the spin/statistics of the particles has not come
into play. We are now interested in describing the ground state of 
a system of $N$ non-interacting, identical fermions and the 
resulting many-body correlations. Such a zero-temperature state 
is obtained by placing all particles in the lowest available 
momentum state, consistent with the Pauli exclusion principle. 
Using fermionic creation and annihilation operators satisfying 
the following anti-commutation relations~\cite{Fetter:1971},
\begin{equation}
 \Big\{A_{\bf p},A_{{\bf p}'}^{\dagger}\Big\}=\delta_{{\bf p},{\bf p}'}
 \;\; {\rm and} \;\Big\{A_{\bf p},A_{{\bf p}'}\Big\}=
 \Big\{A_{\bf p}^{\dagger},A_{{\bf p}'}^{\dagger}\Big\}=0\;,
 \label{FermiOperators}
\end{equation}
the ground state of the system may be written as follows:
\begin{equation}
 |\Phi_{\rm FG}\rangle =  \prod_{{\bf p}=0}^{{\bf p}_{\rm F}}
 A_{\bf p}^{\dagger}|\Phi_{\rm vac}\rangle \;,
 \label{FermiGasWF}
\end{equation}
where $|\Phi_{\rm vac}\rangle$ represents the (non-interacting) vacuum
state and the Fermi momentum ${\bf p}_{\rm F}$ denotes the momentum of
the last occupied single-particle state. Note that henceforth, no
intrinsic quantum number (such as spin and/or isospin) will be
considered. In essence, one assumes that all intrinsic degrees of
freedom have been ``frozen'', thereby concentrating on a single
fermionic species (such as neutrons with spin up). In what follows, 
we compute expectation values of various quantities in the Fermi gas
ground state ($|\Phi_{\rm vac}\rangle$).

We start by computing the Fermi momentum ${\bf p}_{\rm F}$ in terms of 
the number density of the system $\rho\!=\!N/V$. That is,
\begin{equation}
 N = \sum_{{\bf n}} n_{\rm FD}({\bf n}) 
 \mathop{\longrightarrow}_{V\rightarrow\infty}
 V\int\frac{d^{3}p}{(2\pi)^{3}}n_{\rm FD}({\bf p})
 \mathop{=}_{T=0}
 V\frac{p_{\rm F}^{3}}{6\pi^{2}}\;,
 \label{FermiMom1}
\end{equation}
or equivalently,
\begin{equation}
 p_{\rm F}=\left(6\pi^{2}\rho\right)^{1/3} \;.
 \label{FermiMom2}
\end{equation}
Note that in Eq.~(\ref{FermiMom1}) $n_{\rm FD}({\bf p})$ denotes the
Fermi-Dirac occupancy of the single-particle state denoted by 
${\bf p}$ (or ${\bf n}$) and the thermodynamic limit has been assumed.  
As all ground-state observables will be computed over a 
spherically-symmetric Fermi sphere, we define the Fermi-Dirac momentum
distribution $f(q)$ as follows:
\begin{equation}
 f(q)=3q^{2}n_{\rm FD}(q)\;,
 \;\;{\rm with}\;
 \int_{0}^{\infty}f(q)dq=1 \;,
 \label{MomDist}
\end{equation}
where the dimensionless quantity $q\!\equiv\!p/p_{\rm F}$ is the
momentum of the particle in units of the Fermi momentum.

All classical simulations performed and reported in the next sections
must be carried out by necessity at finite temperature. Thus, we now
incorporate finite temperature corrections to the various observables 
of interest. For temperatures $T$ that are small relative to the Fermi 
temperature $T_{\rm F}$ (with $T_{\rm F}\!\equiv\!\epsilon_{\rm F}$), 
finite-temperature corrections may be implemented by means of a 
Sommerfeld expansion~\cite{Pathria:1996}. For example, to lowest order 
in $\tau\!\equiv\!T/T_{\rm F}$ the momentum distribution becomes
\begin{equation}
 f(q,\tau)=\frac{3q^{2}}{\exp{\displaystyle\left[
 {\left(q^{2}-1+\frac{\pi^{2}}{12}\tau^{2}\right)\Big/\tau}\right]+1}} 
 \,\mathop{\longrightarrow}_{\tau\rightarrow0}\
 3q^{2}\Theta(1-q)\;.
 \label{MomDist2}
\end{equation}
Here $\Theta(x)$ is the {\sl ``Heaviside step function''} appropriate
for a zero-temperature Fermi gas. Similarly, the energy-per-particle 
of a {\sl ``cold''} Fermi gas may be readily computed. One 
obtains~\cite{Pathria:1996}
\begin{equation}
 E/N = \epsilon_{\rm F}\int_{0}^{\infty}q^{2}f(q)dq
     = \frac{3}{5}\epsilon_{\rm F}
       \left[1+\frac{5\pi^{2}}{12}\tau^{2}+{\cal O}(\tau^{4})\right]\;,
 \label{FermiEnergy}
\end{equation}
where the Fermi energy is defined by $\epsilon_{\rm F}\!=\!p_{\rm F}^{2}/2m$.

The last Fermi-gas observable that we focus on is the 
{\sl ``two-body correlation function''} $g(r)$. This observable 
measures the probability of finding two particles at a fixed 
distance $r$ from each other. Moreover, the two-body correlation 
function is a fundamental quantity whose Fourier transform yields 
the static structure factor, an observable that may be directly 
extracted from experiment. As such, the two-body correlation 
function is the natural meeting place of theory, experiment, and 
computer simulations~\cite{Vesely:2001}. The two-body correlation 
function may be derived from the density-density correlation 
function~\cite{Fetter:1971}. That is,
\begin{equation}
  g(r) = \frac{1}{\rho^{2}}\rho_{2}({\bf x},{\bf y})
       = \frac{1}{\rho^{2}}\Big\langle 
         \hat{\psi}^{\dagger}({\bf x})\hat{\psi}^{\dagger}({\bf y})
         \hat{\psi}({\bf y})\hat{\psi}({\bf x})\Big\rangle \;,
 \label{TwoBodyCorr}
\end{equation}
where $\hat{\psi}({\bf x})$ is a fermionic field operator and 
the Dirac brackets denote a thermal expectation value. As the 
two-body correlation function for a non-interacting Fermi gas 
may be readily evaluated at zero temperature~\cite{Fetter:1971}, 
we only provide its extension to finite temperatures. To lowest
order in $\tau\!\equiv\!T/T_{\rm F}$ (and for a single fermionic 
species) one obtains
\begin{equation}
  g(r) = 1 - \left(3\,\frac{j_{1}(z)}{z}\right)^{2}
         \left[1-\frac{\pi^{2}}{12}z^{2}\tau^{2}\right]\;,
 \label{TwoBodyCorrFG}
\end{equation}
where $z\!\equiv\!p_{\rm F}r$ and $j_{1}(z)$ is the spherical Bessel
function of order $n\!=\!1$, namely,
\begin{equation}
  j_{1}(z) = \frac{\sin(z)}{z^{2}}-\frac{\cos(z)}{z} \;.
 \label{SphericalBessel1}
\end{equation}

Equations~(\ref{MomDist2}), (\ref{FermiEnergy}), and (\ref{TwoBodyCorrFG})
display the three fundamental observables of a cold Fermi gas that we
aim to reproduce in this work via a momentum-dependent, two-body Pauli
potential. Note that in most (if not all) earlier studies of this
kind, only the kinetic energy of the Fermi gas
[Eq.~(\ref{FermiEnergy})] was used to constrain the parameters of the
Pauli potential~\cite{Wilets:1977,Dorso:1987,Boal:1988,Peilert:1992,
Maruyama:1997rp,Watanabe:2003}. We are unaware of any earlier effort
at including more sensitive Fermi-gas observables to constrain the
parameters of the model. Clearly, it should be possible to reproduce 
the second moment of the distribution ({\it i.e.,} the kinetic energy)
even with an incorrect momentum distribution.  Thus, while we build on
earlier approaches, we also highlight some of their shortcomings.

\subsection{Pauli Potential: A New Functional Form}
\label{sec:PauliPotential}

In the previous section the wave function of a zero-temperature Fermi 
gas was introduced as follows:
\begin{equation}
 |\Phi_{\rm FG}\rangle =  \prod_{{\bf p}=0}^{{\bf p}_{\rm F}}
 A_{\bf p}^{\dagger}|\Phi_{\rm vac}\rangle \;.
 \label{FermiGasWF2}
\end{equation}
Essential to the dynamical behavior of the system are the anti-commutation
relations [Eq.~(\ref{FermiOperators})] that enforce the Pauli exclusion 
principle ({\it i.e.,} $(A_{\bf p}^{\dagger})^{2}\!\equiv\!0$). As it is often 
done, one may project the above ``second-quantized'' form of the many-fermion 
wave-function into configuration space to obtain the well-known Slater 
determinant. That is,
\begin{equation}
  \Phi_{FG}({\bf p}_{1},\dots,{\bf p}_{N};
           {\bf r}_{1},\dots,{\bf r}_{N})=\frac{1}{\sqrt{N!}}
   \left|\begin{matrix}
      \varphi_{{\bf p}_{1}}({\bf r}_{1}) & \ldots & 
      \varphi_{{\bf p}_{1}}({\bf r}_{N})   \\
      \vspace{-0.1cm} . & \ldots &  .    \\ 
      \vspace{-0.1cm} . & \ldots &  .    \\ 
      \vspace{-0.1cm} . & \ldots &  .    \\ 
      \vspace{+0.1cm}
      \varphi_{{\bf p}_{N}}({\bf r}_{1}) & \ldots & 
      \varphi_{{\bf p}_{N}}({\bf r}_{N}) 
    \end{matrix} \right|\;,
 \label{Slater}
\end{equation}
where the single-particle wave-functions $\varphi_{\bf p}({\bf r})$
are the (``box'') normalized plane waves defined in Eq.~(\ref{FreePhi}). 
The Slater determinant embodies important correlations that were discussed 
in the previous section and that we aim to incorporate into our classical 
simulations. These are:
\begin{description}
 \item{(a)} At zero temperature, only momentum states having a 
            magnitude $|{\bf p}|$ less than the Fermi momentum 
            $p_{\rm F}$ are occupied; the rest are empty.
            \vspace{-0.20cm} 
 \item{(b)} As a consequence of the Pauli exclusion principle
            [$(A_{\bf p}^{\dagger})^{2}\!\equiv\!0$], the probability
            that two fermions share the same identical momentum is 
            equal to zero. Mathematically, this result follows from 
            the fact that the wave-function vanishes whenever two 
            rows of the Slater determinant are equal to each other.
            \vspace{-0.20cm} 
 \item{(c)} Similarly, the wave-function also vanishes whenever two 
            columns of the Slater determinant are identical. This
            fact precludes two fermions from occupying the same exact 
            location in space.
            \vspace{-0.20cm} 
\end{description}
The first two properties are embedded in the momentum distribution 
of Eq.~(\ref{MomDist2}), namely, a quadratic momentum distribution
sharply peaked at the Fermi momentum $p_{\rm F}$ (recall that such a
momentum distribution emerges after folding the Heaviside step
function with the phase space factor). The third property induces
spatial correlations that are captured by the two-body correlation
function $g(r)$ of Eq.~(\ref{TwoBodyCorrFG}). Indeed, the two-body
correlation function is related to the integral of the Slater
determinant over all but two of the coordinates of the particles 
({\it e.g.,} ${\bf r}_{1}$ and ${\bf r}_{2}$). Clearly, the Slater 
determinant vanishes whenever ${\bf r}_{1}={\bf r}_{2}$, and so does 
the two-body correlation function at 
$r\!=\!|{\bf r}_{1}-{\bf r}_{2}|\!\equiv\!0$. It is the aim of this 
contribution to build a Pauli potential that incorporates these three 
fundamental properties of a free Fermi gas.

However, before doing so, we briefly review the Pauli potential
introduced by Wilets and collaborators --- and used by others with
minor modifications --- to simulate the collisions of heavy ions and
the properties of neutron rich matter at sub-saturation densities. 
Such a Pauli potential is given by a sum of momentum-dependent, 
two-body terms of the following form:
%%%
\begin{equation}
 V_{\rm Pauli}({\bf p}_{1},\dots,{\bf p}_{N};
               {\bf r}_{1},\dots,{\bf r}_{N})=
               \sum_{i<j=1}^{N} V_{0}\exp(-s_{ij}^{2}/2) \;,
 \label{VPauli0}
\end{equation}
%%%
where $V_{0}\!>\!0$ and the dimensionless phase-space {\sl ``distance''} 
between points $({\bf p}_{i},{\bf r}_{i})$ and $({\bf p}_{j},{\bf r}_{j})$ 
is given by
%%%
\begin{equation}
  s_{ij}^{2}\equiv 
  \frac{|{\bf p}_{i}-{\bf p}_{j}|^{2}}{p_{0}^{2}} +
  \frac{|{\bf r}_{i}-{\bf r}_{j}|^{2}}{r_{0}^{2}} \;.
 \label{sij}
\end{equation}
%%%
Here $p_{0}$ and $r_{0}$ are momentum and length scales related to the
excluded phase-space volume that is used to mimic fermionic correlations. 
That is, whenever the phase-space distance between
two particles is such that $s_{ij}^{2}\!\lesssim\!1$, then a penalty
is levied on the system in an effort to mimic the Pauli exclusion
principle. Although the parameters of this Pauli potential ($V_{0}$,
$p_{0}$, and $r_{0}$) can --- and have --- been adjusted to reproduce
the kinetic energy of a free Fermi gas, it fails (as we show later)
in reproducing more sensitive Fermi-gas observables, particularly,
the momentum distribution $f(p)$ and the two-body correlation function
$g(r)$.

Upon closer examination, the above flaws should not come as a
surprise.  In our previous discussion of the Slater determinant it has
been established that the probability of finding two identical
fermions in the same location in space {\sl ``or''} with the same
momenta must be identically equal to zero. Yet the Pauli potential of
Eq.~(\ref{VPauli0}) fails to incorporate this important dynamical
behavior.  Indeed, the above Pauli potential imposes a penalty on the
system only when both the location {\sl ``and''} momenta of the two
particles are close to each other ({\it i.e.,}
$s_{ij}^{2}\!\lesssim\!1$).  In particular, no penalty is imposed
whenever two fermions occupy the same location in space, provided that
their momenta are significantly different from each other, {\it i.e.,}
$|{\bf p}_{i}-{\bf p}_{j}|^{2}\!\gg\!p_{0}^{2}$.  Thus, the Pauli
potential of Eq.~(\ref{VPauli0}) will generate an incorrect two-body
correlation function, namely, one with $g(r)\!\ne\!0$ as r tends to
zero. By the same token, an incorrect momentum distribution will be
generated, although not necessarily its second moment.

To remedy these deficiencies, a Pauli potential is now constructed so 
that the three properties [(a), (b), and (c)] defined above are explicitly 
satisfied. To this end, we introduce the following form for the Pauli 
potential:
%%%
\begin{widetext}
\begin{eqnarray}
 V_{\rm Pauli}({\bf p}_{1},\dots,{\bf p}_{N};
               {\bf r}_{1},\dots,{\bf r}_{N}) 
    &=&  \sum_{i<j=1}^{N} \Big[V_{A}\exp(-r_{ij}/r_{0})  
                            +V_{B}\exp(-p_{ij}/p_{0})\Big] 
    \nonumber \\ 
    &+& \sum_{i=1}^{N} V_{C}\,\Theta_{\eta}(q_{i})\;, 
\label{VPauli1}
\end{eqnarray}
\end{widetext}
where $r_{ij}\!=\!|{\bf r}_{i}-{\bf r}_{j}|$,
      $p_{ij}\!=\!|{\bf p}_{i}-{\bf p}_{j}|$,
      $q_{i}\!=\!|{\bf p}_{i}|/p_{\rm F}$,
and $\Theta_{\eta}$ is a suitably smeared Heaviside-step function of the 
following form:
%%%
\begin{equation}
  \Theta_{\eta}(q)\equiv \frac{1}{1+\exp[-\eta(q^{2}-1)]}
  \mathop{\longrightarrow}_{\eta\rightarrow\infty}
  \Theta(q)\;.
 \label{SmearedTheta}
\end{equation}
%%%
The parameters of the model $V_{A},V_{B},V_{C}$ and $r_{0},p_{0},\eta$
will be adjusted to reproduce both the momentum distribution and
two-body correlation function of a low-temperature Fermi gas.  Note
that the phase-space dependence of the Pauli potential has been
separated into a {\sl ``sum''} of two-body pieces, with the first
acting exclusively in configuration space and the second one only in
momentum space.  The first term in the potential imposes a penalty as
the particles get too close (of the order of $r_{0}$) to each
other. Similarly, the second term in the potential penalizes particles
whenever their relative momenta becomes of the order of
$p_{0}$. Finally, the third {\sl ``one-body''} term enforces the
low-temperature behavior of the Fermi gas, namely, that the
probability of finding any particle with a momentum significantly
larger than the Fermi momentum is vanishingly small.  Most of the
parameters will depend explicitly on the density of the system (see
Sec.~\ref{sec:results}).  This reflects the complex many-body
nature of the Pauli correlations and our inability to simulate them
by means of a {\sl ``simple''} (albeit momentum dependent) two-body 
potential.

\section{Results}
\label{sec:results}
We start this section by listing in Table~\ref{Table1} the parameters
of the Pauli potential. For the strength parameters, the following
simple density dependence is assumed:
%%%
\begin{equation}
  V_{i}(\rho)= V_{i}^{0}\Big(\rho/\rho_{0}\Big)^{\alpha_{i}}\;,
  \quad i=\{A,B,C\} \;,
 \label{Strength}
\end{equation}
%%%
where $\rho_{0}\!=\!\rho_{sat}/4\!=\!0.037~{\rm fm}^{-3}$ is the density
of a single fermionic species (for example, neutrons with spin up)
at nuclear-matter saturation density ($\rho_{sat}$ = 0.148 fm$^{-3}$).
For the range parameters 
($r_{0}$ and $p_{0}$) the following scaling relation is adopted:
%%%
\begin{subequations}
\begin{align}
  & r_{0}=\beta_{A}/p_{\rm F}  \;,\\
  & p_{0}=\beta_{B}p_{\rm F}   \;.
\end{align}
\label{Ranges}
\end{subequations}
%%%

%%%%%%%%%%%%%%%%%%%%%%%%%%%%%%%%%%%%%%%%%%%%%%%%%%%%%%%%%%%%%%%%%
\begin{table}
\begin{tabular}{|c|c|c|c|c|c|c|c|c|}
 \hline
 $V_{A}^{0}$ & $V_{B}^{0}$ & $V_{C}^{0}$ & $\alpha_{A}$ & 
 $\alpha_{B}$ & $\alpha_{C}$ & $\beta_{A}$ & $\beta_{B}$ & $\eta$ \\
 \hline
 \hline
  13.517 & 1.260 & 3.560 & 0.629 & 0.665 & 
   0.831 & 0.845 & 0.193 & 30         \\
\hline
\end{tabular}
\caption{Strength (in MeV) and range parameters (dimensionless)
         for the various components of the Pauli potential. See 
         Eqs.~(\ref{VPauli1}), (\ref{Strength}), and~(\ref{Ranges}).
         These values have been used to simulate a system of 
         $N\!=\!1000$ identical fermions at a temperature of 
         $\tau\!=\!T/T_{\rm F}\!=\!0.05$.}
\label{Table1}
\end{table}
%%%%%%%%%%%%%%%%%%%%%%%%%%%%%%%%%%%%%%%%%%%%%%%%%%%%%%%%%%%%%%%%%

Monte-Carlo simulations for a system of $N\!=\!1000$ identical
fermions at the finite (but small) temperature of $\tau\!=\!T/T_{\rm
F}\!=\!0.05$ have been performed. Initially, the particles are
distributed randomly throughout the box with momenta that are
uniformly distributed up to a maximum momentum of the order of
the Fermi momentum. After an initial thermalization phase of typically
2000 sweeps (or 2 million Monte Carlo moves for both coordinates and
momenta) data is accumulated for an additional 2000 sweeps, with the
data divided into 10 groups to avoid correlations among the data. It
is from these 10 groups that averages and errors are generated.

\subsection{Kinetic Energy}
\label{sec:KinEn}

\begin{figure}[ht]
\vspace{0.50in}
\includegraphics[width=4.5in,angle=0]{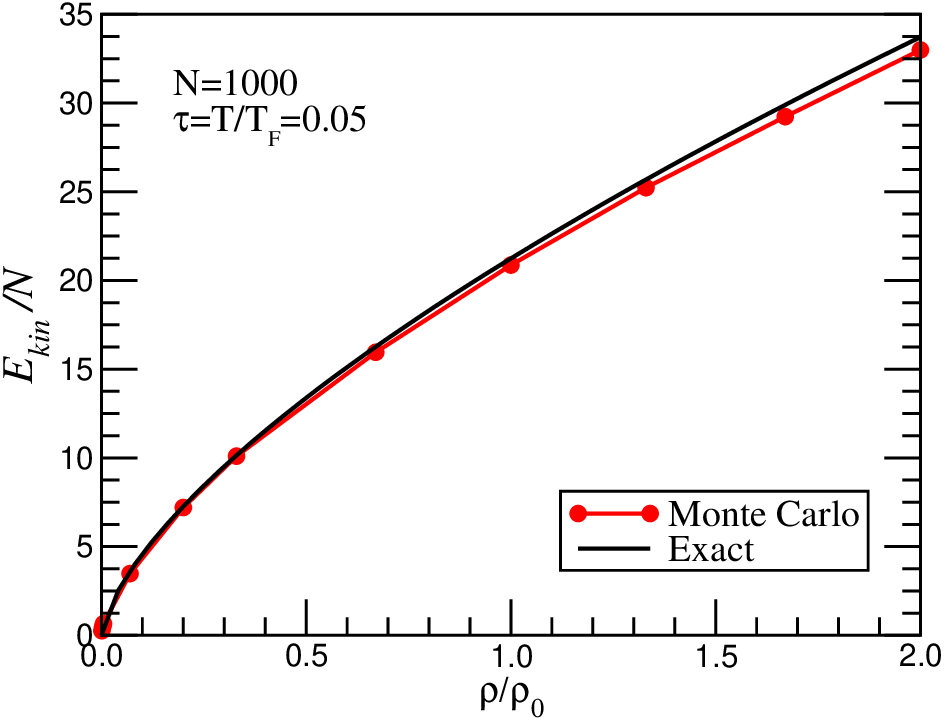}
\caption{(color online) Average kinetic energy of a system of
         $N\!=\!1000$ identical fermions at a temperature of 
         $\tau\!=\!T/T_{\rm F}\!=\!0.05$. The line with circles
         is the result of the Monte Carlo simulations with the
         Pauli potential of Eq.~(\ref{VPauli1}). The solid line
         is the exact behavior of a non-relativistic Fermi gas,
         as given by Eq.~(\ref{FermiEnergy}).}
\label{Fig1}
\end{figure}

The kinetic energy of a finite-temperature Fermi gas as a function
of density is displayed in Fig.\ref{Fig1}. The (black) solid line 
represents the analytic behavior of a free Fermi gas correct to 
second order in $\tau$. The (red) line with circles is the result 
of the Monte Carlo simulations with the Pauli potential defined in 
Eq.~(\ref{VPauli1}). The agreement (to better than 5\%) is as good 
as the one obtained with earlier parametrizations of the Pauli 
potential.

%%%%%%%%%%%%%%%%%%%%%%%%%%%%%%%%%%%%%%%%%%%%%%%%%%%%%%%%%%%%%%%%%%%%%%
%%%%%%%%%%%%%%%%%%%%%%%%%%%%%%%%%%%%%%%%%%%%%%%%%%%%%%%%%%%%%%%%%%%%%%

To our knowledge, reproducing the kinetic energy of a free Fermi gas
is the sole constraint that has been imposed on most of the Pauli
potentials available to date. However, while a host of Pauli
potentials can reproduce such a behavior, it is unclear if these
potentials can also reproduce the full momentum distribution.
Thus, we now show how the Pauli potential defined in 
Eq.~(\ref{VPauli1}) is successful at reproducing two highly sensitive
Fermi-gas observables, namely, the momentum distribution and 
the two-body correlation function.

\subsection{Momentum Distribution}
\label{sec:MomDis}

\begin{figure}[ht]
\vspace{0.50in}
\includegraphics[width=4.5in,angle=0]{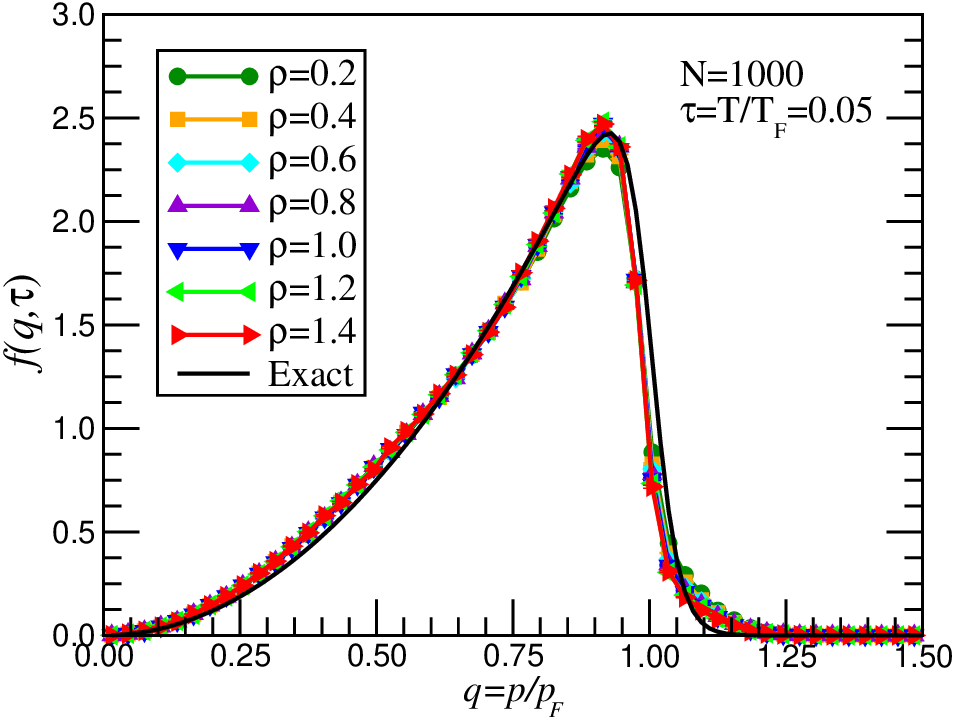}
\caption{(color online) Momentum distribution of a system of
         $N\!=\!1000$ identical fermions at a temperature of 
         $\tau\!=\!T/T_{\rm F}\!=\!0.05$ for a variety of
         densities (expressed in units of 
         $\rho_{0}\!=\!0.037~{\rm fm}^{-3}$). The momentum 
         distribution has been normalized  to one [see 
         Eq.~(\ref{MomDist})]. The (black) solid line
         with no symbols gives the exact behavior of 
         a non-relativistic Fermi gas.}
\label{Fig2}
\end{figure}

The momentum distribution obtained from the Monte-Carlo simulations is 
displayed in Fig.~\ref{Fig2} for a variety of densities. Note that the 
momentum distribution has been normalized to one and that the densities 
have been expressed in units of $\rho_{0}\!=\!0.037~{\rm fm}^{-3}$. As 
indicated in Eq.~(\ref{MomDist2}), the momentum distribution of a cold 
Fermi gas depends solely on the two dimensionless ratios 
$q\!=\!p/p_{\rm F}$ and $\tau\!=\!T/T_{\rm F}$. Thus, all curves must 
collapse into the exact one --- displayed by the (black) solid line --- 
independent of density. It is gratifying to see that this is indeed the
case. In contrast, we show in the next section that the {\sl standard} 
Pauli potential of Eq.~(\ref{VPauli0}) fails to reproduce this behavior.

%%%%%%%%%%%%%%%%%%%%%%%%%%%%%%%%%%%%%%%%%%%%%%%%%%%%%%%%%%%%%%%%%%%%%%
%%%%%%%%%%%%%%%%%%%%%%%%%%%%%%%%%%%%%%%%%%%%%%%%%%%%%%%%%%%%%%%%%%%%%%

\subsection{Two-Body Correlation Function}
\label{sec:TwoBody}

\begin{figure}[ht]
\vspace{0.50in} \includegraphics[width=4.5in,angle=0]{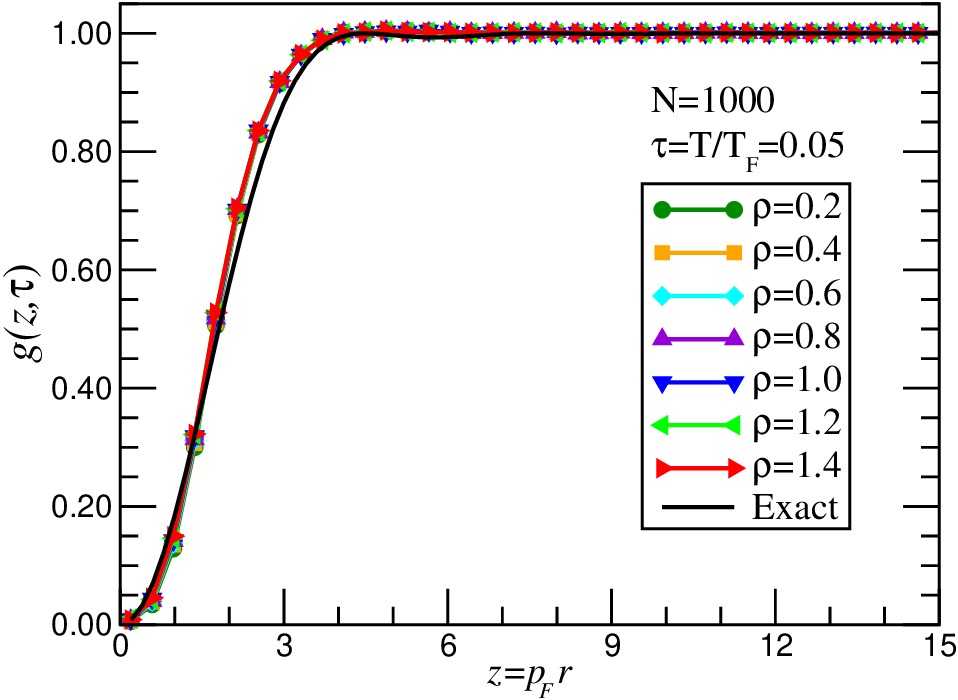}
\caption{(color online) Two-body correlation function for a 
         system of $N\!=\!1000$ identical fermions at a 
         temperature of $\tau\!=\!T/T_{\rm F}\!=\!0.05$ for 
         a variety of densities (expressed in units of 
         $\rho_{0}\!=\!0.037~{\rm fm}^{-3}$). The (black) 
         solid line with no symbols gives the exact behavior 
         of a non-relativistic Fermi gas.}
\label{Fig3}
\end{figure}

The two-body correlation function of a free Fermi gas is displayed in
Fig.~\ref{Fig3}. The two-body correlation function $g(r)$ is related
to the probability of finding two particles separated by a distance
$r$.  For a free Fermi gas, the probability of finding two identical
fermions at zero separation is identically equal to zero [see
Eq.~(\ref{TwoBodyCorrFG})]. As this short-range (anti-)correlation is
the sole consequence of the Pauli exclusion principle, the 
{\sl ``Pauli hole''} disappears for distances of the order of the
interparticle separation ($p_{\rm F}^{-1}$). As in the case of the
momentum distribution, the correlation function depends only on 
two dimensionless variables ($z\!=\!p_{\rm F}r$ and $\tau$). It is
again gratifying that the simulation curves scale to the exact correlation 
function, depicted here with a (black) solid line.

%%%%%%%%%%%%%%%%%%%%%%%%%%%%%%%%%%%%%%%%%%%%%%%%%%%%%%%%%%%%%%%%%%%%%%
%%%%%%%%%%%%%%%%%%%%%%%%%%%%%%%%%%%%%%%%%%%%%%%%%%%%%%%%%%%%%%%%%%%%%%

\subsection{Comparison to other approaches}
\label{sec:Comparison}

In this section we compare the Pauli potential introduced in
Eq.~(\ref{VPauli1}) to earlier approaches that are based on
Eq.~(\ref{VPauli0}). Such approaches have been very successful 
in reproducing the kinetic energy of a free Fermi gas for a wide 
range of densities. Indeed, the kinetic energy displayed in Fig.~\ref{Fig2}
of Ref.~\cite{Maruyama:1997rp} is as good --- if not better --- 
than the one obtained here.
%%%%%%%%%%%%%%%%%%%%%%%%%%%%%%%%%%%%%%%%%%%%%%%%%%%%%%%%%%%%%%%%%%%%%%
\begin{figure}[ht]
\vspace{0.50in}
\includegraphics[width=5.5in,angle=0]{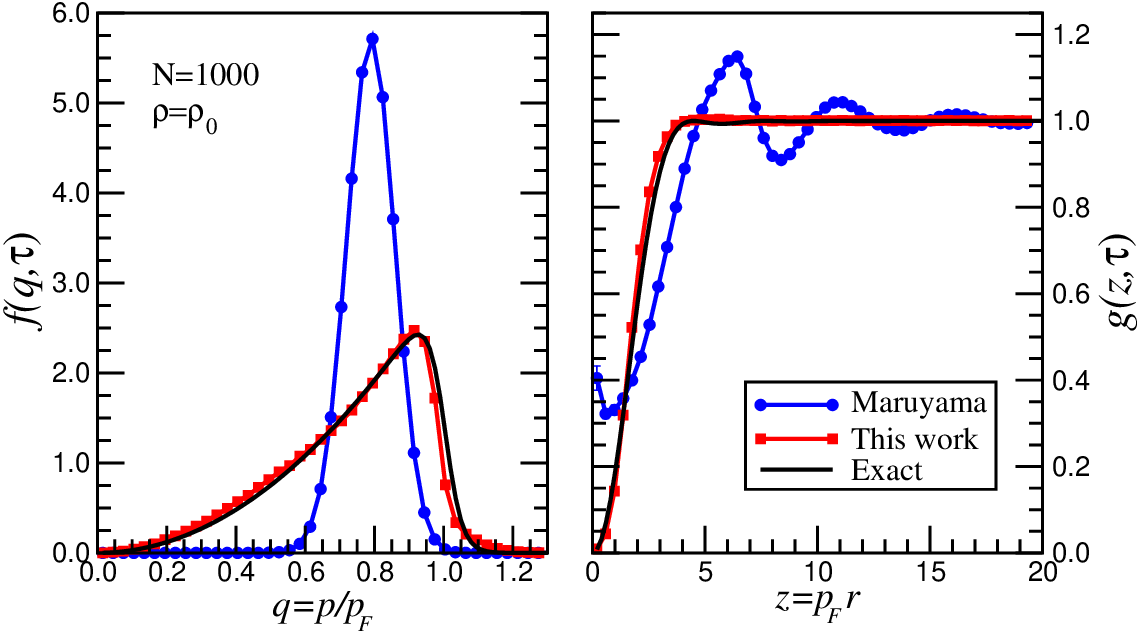}
\caption{(color online) Comparison between the Pauli potential
         introduced in this work [Eq.~(\ref{VPauli1})] and 
         earlier approaches based on Eq.~(\ref{VPauli0}). The 
         left-hand panel shows the momentum distribution while 
         the right-hand panel the two-body correlation function.
         The simulations have been carried out for a system of 
         $N\!=\!1000$ identical fermions at a density of
         $\rho_{0}\!=\!0.037~{\rm fm}^{-3}$.
         The (blue) line with circles is obtained from a Monte
         Carlo simulation using Eq.~(\ref{VPauli0}) with the 
         parameters of Ref.~\cite{Maruyama:1997rp}. The (red) 
         line with squares displays the results using the 
         Pauli potential introduced in this work; the black line 
         gives the exact behavior of a non-relativistic Fermi gas.}
\label{Fig4}
\end{figure}
%%%%%%%%%%%%%%%%%%%%%%%%%%%%%%%%%%%%%%%%%%%%%%%%%%%%%%%%%%%%%%%%%%%%%%
However, a faithful reproduction of the kinetic energy does not
guarantee that the system displays the same phase-space correlations
as that of a free Fermi gas. To illustrate this point we compare in
Fig.~\ref{Fig4} the results from the two approaches for the momentum
distribution and two-body correlation function at a fixed density of
$\rho_{0}\!=\!0.037~{\rm fm}^{-3}$. The left-hand panel in the figure
displays the momentum distribution and indicates that while earlier
approaches (depicted with a blue line with circles) are accurate at
reproducing the second moment of the distribution ({\it i.e.,} the
kinetic energy) the momentum distribution itself shows a behavior that
differs significantly from that of a cold Fermi gas. The right-hand
panel shows deficiencies that are as --- or even more --- severe. Two 
points are worth highlighting. First, the two-body correlation
function $g(r)$ generated with the standard form of the Pauli
potential does not vanish at $r\!=\!0$. Second, for distances of the
order of the inter-particle separation and beyond, the two-body
correlation function develops artificial oscillations. Failure in
reproducing the correct behavior of $g(r)$ at $r\!=\!0$ is relatively
simple to understand. Potentials based on Eq.~(\ref{VPauli0}) impose a
penalty on the system only if {\sl both} the positions and momenta of
the two fermions are close to each other. Yet the correlations embodied 
in a free Fermi gas are significantly more stringent than that. Indeed, 
a Slater determinant vanishes if {\sl either} the positions or the momenta 
of the two fermions are equal to each other. In contrast, the development 
of artificial structure in $g(r)$ is a more subtle effect that is intimately 
related to the momentum dependence of the potential and that we now address.

The artificial oscillations present in $g(r)$ are reminiscent of the
structure of liquids and/or crystals. The emergence of crystalline
structure in a low-temperature/low-density system would be expected if
the energy of localization becomes small relative to the mutual
repulsion between the particles. Such, however, is not the behavior of
a cold Fermi gas. While spatial anti-correlations would favor the
formation of a periodic structure, the large velocities of the
particles --- resulting from the Pauli exclusion principle --- would
make their localization extremely costly. What is not evident is the
reason for the momentum distribution generated with a standard Pauli
potential to lead to {\sl crystallization}, whereas not that of a real
Fermi gas (see left-hand panel in Fig.~\ref{Fig4}).

To our knowledge, the answer to this question was first provided by 
Neumann and Fai~\cite{Neumann:1994bn}. One must realize that any 
Hamiltonian that contains momentum-dependent interactions --- such 
as most (if not all) Pauli potentials --- generates a {\sl ``canonical''} 
momentum distribution that may (and does!) differ significantly from the
corresponding {\sl ``kinematical''} momentum distribution 
$\langle{\bm\pi}\rangle\!\equiv\!\langle m\dot{{\bf r}}\rangle$. 
Indeed, in a Hamiltonian formalism where the Hamiltonian depends
on the positions and canonical (not kinematical) momenta of all
the particles, namely, $H = H({\bf p}_{1},\dots,{\bf p}_{N};
                              {\bf r}_{1},\dots,{\bf r}_{N})$
the kinematical velocities must be obtained from Hamilton's equations
of motion, {\it i.e.,} $\dot{\bf r}_{i}={\partial H}/\partial {\bf p}_{i}$.
For a Hamiltonian that contains momentum-dependent interactions --- as 
in the case of the Pauli potential --- then the kinematical momentum 
${\bm\pi}_{i}\!\equiv\!m\dot{{\bf r}}_{i}$ differs from the 
corresponding canonical momentum ${\bf p}_{i}$. This suggests 
that while a given choice of Pauli potential may produce the correct 
{\sl canonical} momentum distribution, it may generate kinematical
velocities that are too small to prevent crystallization. Such a
distinction between canonical and kinematical momenta is an unwelcome,
yet unavoidable, consequence of the approach.  A cold Fermi gas --- a
system subjected to no interactions --- is the quintessential quantum
system were such an artificial distinction is not required.

%%%%%%%%%%%%%%%%%%%%%%%%%%%%%%%%%%%%%%%%%%%%%%%%%%%%%%%%%%%%%%%%%%%%%%
\begin{figure}[ht]
\vspace{0.50in}
\includegraphics[width=4.5in,angle=0]{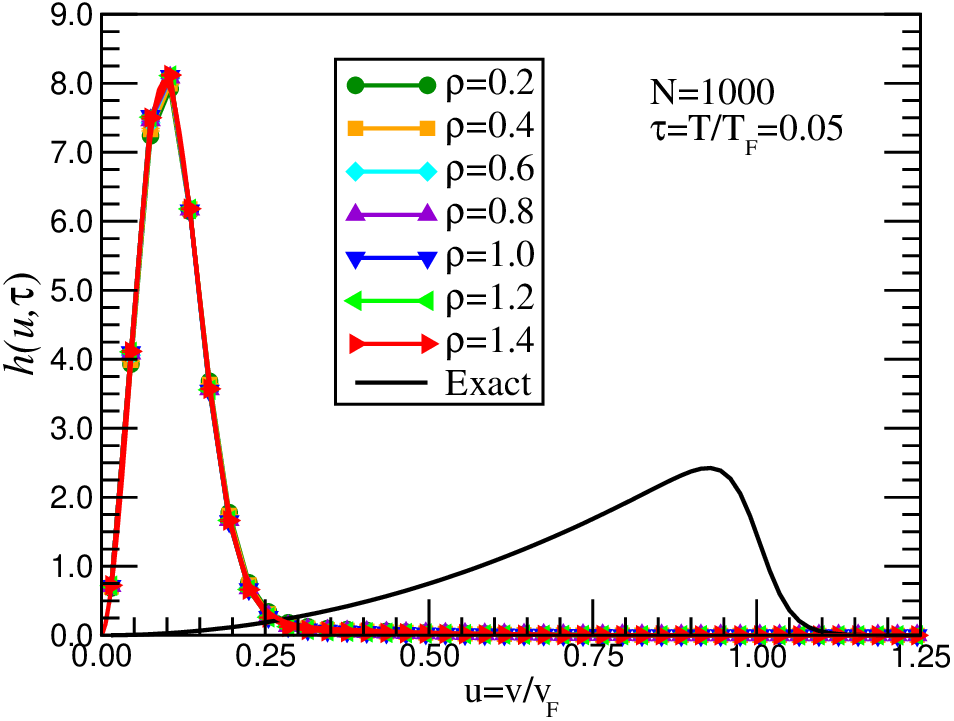}
\caption{(color online) {\sl ``Kinematical''} velocity  
         distribution of a system of $N\!=\!1000$ identical 
         fermions at a temperature of 
         $\tau\!=\!T/T_{\rm F}\!=\!0.05$ for a variety of
         densities (expressed in units of 
         $\rho_{0}\!=\!0.037~{\rm fm}^{-3}$). The 
         velocity distribution has been normalized to one.
         The (black) solid line with no symbols gives the 
         exact behavior of a non-relativistic Fermi gas.}
\label{Fig5}
\end{figure}
%%%%%%%%%%%%%%%%%%%%%%%%%%%%%%%%%%%%%%%%%%%%%%%%%%%%%%%%%%%%%%%%%%%%%%

It is worth noting that although some earlier choices yield an extremely 
soft kinematical momentum distribution, the Pauli potential introduced 
in this work is not immune to such a disease. Indeed, we believe that
an unrealistically soft kinematical momentum distribution is likely to 
be a general result of the Hamiltonian approach. Thus, we close this 
section by displaying in Fig.~\ref{Fig5} the kinematical {\sl velocity} 
distribution obtained with the new choice of Pauli potential introduced 
in Eq.~(\ref{VPauli1}). For comparison, the exact distribution of a cold
Fermi gas (solid black line) is also included. The dependence of the
Pauli potential on the canonical momentum distribution is responsible
for generating such a soft velocity distribution, with its peak around
1/10 of the Fermi velocity. And while we were able to avoid
crystallization with the present set of parameters (see
Figs.~\ref{Fig3} and~\ref{Fig4}), the risk of crystallization looms
large (see next section).
%%%%%%%%%%%%%%%%%%%%%%%%%%%%%%%%%%%%%%%%%%%%%%%%%%%%%%%%%%%%%%%%%%%%%%

\subsection{Finite-Size Effects}
\label{sec:Caution}

We have observed that the Pauli potential introduced in
Eq.~(\ref{VPauli1}), with its parameters suitably adjusted, has been
successful in reproducing a variety of Fermi-gas observables, such as
its kinetic energy, its (canonical) momentum distribution, and its
two-body correlation function.  However, the first indication of a
potential problem --- and one that may be generic to all approaches
employing momentum-dependent interactions --- is the emergence of an
unrealistically soft velocity distribution and with it, the
possibility of artificial spatial correlations ({\it i.e.,}
crystallization).  Fortunately, with the choice of parameters adopted
in this work, the problem of crystallization was avoided (see
Fig.~\ref{Fig3}). Yet, there is no guarantee that crystallization will
not become a problem as one examines the sensitivity of our results to
{\sl finite-size effects}.

%%%%%%%%%%%%%%%%%%%%%%%%%%%%%%%%%%%%%%%%%%%%%%%%%%%%%%%%%%%%%%%%%%%%%%
\begin{figure}[ht]
\vspace{0.50in}
\includegraphics[width=4.5in,angle=0]{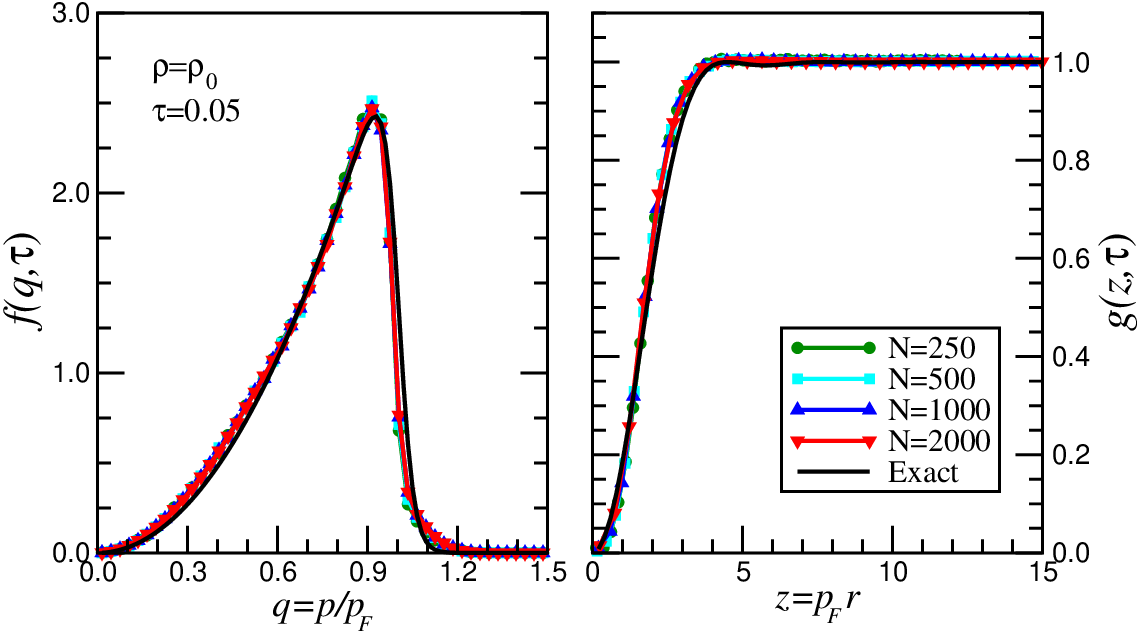}
\caption{(color online) Finite-size effects on the canonical
         momentum distribution (left-hand panel) and the 
         two-body correlation function (right-hand panel) 
         for a system of identical fermions at a temperature 
         of $\tau\!=\!T/T_{\rm F}\!=\!0.05$ and a density 
         of $\rho =\!\rho_{0}\!=\!0.037~{\rm fm}^{-3}$. 
         Simulations were carried out for systems containing
         $N\!=\!250, 500, 1000$ and $2000$ particles. The 
         (black) solid line with no symbols gives the exact 
         behavior of a non-relativistic Fermi gas.}
\label{Fig6}
\end{figure}
%%%%%%%%%%%%%%%%%%%%%%%%%%%%%%%%%%%%%%%%%%%%%%%%%%%%%%%%%%%%%%%%%%%%%%

In order to estimate the sensitivity of our results to finite-size
effects, Monte Carlo simulations were performed for a system
containing $N\!=\!250$, $N\!=\!500$, $N\!=\!1000$, and $N\!=\!2000$,
identical fermions (note that the results reported so far have been
limited to 1000 particles). The conclusions from this study are mixed.
First (and fortunately) no evidence of crystallization or of 
significant finite-size effects were found. These findings are 
displayed in Fig.~\ref{Fig6} for both the canonical momentum 
distribution (left-hand panel) and the two-body correlation 
function (right-hand) panel.
Unfortunately, however, in order to preserve the high quality 
of the results previously obtained with 1000 particles, a parameter 
of the Pauli potential [$V_{B}^{0}$ in Eq.~(\ref{VPauli1})] had 
to be fine tuned. Specifically, the following scaling with particle 
number was used:
%%%
\begin{equation}
 V_{B}^{0}(N) = V_{B}^{0}(N\!=\!1000)\,\left(\frac{1000}{N}\right)\;, 
\end{equation}
%%%
with $V_{B}^{0}(N\!=\!1000)\!=\!1.26$~MeV being the value listed in
Table~\ref{Table1}. This unpalatable fact may be a reflection of 
the highly challenging task at hand: how to simulate fermionic
many-body correlations by means of a ``simple'' two-body Pauli 
potential.

%%%%%%%%%%%%%%%%%%%%%%%%%%%%%%%%%%%%%%%%%%%%%%%%%%%%%%%%%%%%%%%%%%%%%%

\section{Conclusions}
\label{sec:conclusions}

Semi-classical simulations of fermionic systems attempt to circumvent
the many challenges posed by bona-fide quantum simulations through the
inclusion of a fictitious Pauli potential~\cite{Wilets:1977,Dorso:1987,
Boal:1988,Peilert:1992,Maruyama:1997rp,Watanabe:2003}. In the present
contribution we conducted a critical study of the features and
predictions of a widely used version of the Pauli potential. We
concluded that the constraints imposed by such a Pauli potential,
namely, the suppression of phase-space configurations having two
fermions with {\sl both} positions and momenta similar to each other,
are too weak to faithfully reproduce some basic properties of a free
Fermi gas. Instead, by examining the well-known behavior of the Slater
determinant we suggest that phase-space configurations should be
suppressed when {\sl either} the positions or the momenta of the
fermions are close to each other. By incorporating these features into
a new form of the Pauli potential --- and by carefully tuning the
parameters of the model --- the momentum distribution and the two-body
correlation function of a free Fermi gas were accurately reproduced.

In the course of this study a pathology that is generic to all Pauli
potential was uncovered. As pointed out by Neumann and
Fai~\cite{Neumann:1994bn} (to our knowledge for the first time), the
{\sl ``canonical''} momentum distribution generated via Monte Carlo
(or other) methods may differ significantly from the resulting {\sl
``kinematical''} momentum distribution. This suggests that while the
kinetic energy of the free Fermi gas (computed from the canonical
momenta) may be accurately reproduced, the distribution of velocities
may be grossly distorted. Indeed, we found a distribution of
velocities that significantly under-estimates --- by a factor of 10
--- that of a free Fermi gas. Such {\sl ``sluggishness''} among the
particles could have disastrous consequences by inducing artificial
ordering in the system ({\it e.g.,} {\sl ``crystallization''}). The
possible appearance of artificial long-range order in the system must
be examined on a case by case basis. For example, with the standard
version of the Pauli potential~\cite{Maruyama:1997rp} the system 
displays an anomalous two-body correlation function suggestive of 
crystallization. On the other hand, the Pauli potential introduced 
in this work faithfully reproduces the two-body correlation function 
of a free Fermi gas.

In conclusion a systematic study of the standard version of the Pauli
potential has been conducted. While simple and widely used, such a
version fails to reproduce some of the most basic properties of a free
Fermi gas. Moreover, an analysis of the two-body correlation function
generated by such a model reveals artificial long-range order. To
correct these flaws a refined version of the Pauli potential was
introduced. This new version --- inspired by the properties of a
Slater determinant --- generated accurate (canonical) momentum
distribution and two-body correlation functions, while avoiding
crystallization. Yet the {\sl ``kinematical''} momentum distribution
generated from such a Pauli potential was grossly underestimated. We
believe this to be a generic feature of any momentum-dependent Pauli
potential that is used in conjunction with a Hamiltonian
approach~\cite{Neumann:1994bn}. Thus, the possibility for generating
artificial correlations in the system ({\it e.g.,} crystallization)
looms large.

%------------------------------------------------------------------------------
\section{ACKNOWLEDGMENTS}

\smallskip
This work was supported in part by the United States Department of
Energy grant DE- FG05-92ER40750 and by the Spanish Ministry of
education under project DGI-FIS2006- 05319. One of the authors,
M.A.P.G. would like to dedicate this work to the memory of J.M.L.G.

%%%%%%%%%%%%%%%%%%%%%%%%%%%%%%%%%%%%%%%%%%%%%%%%%%%%%%%%%%%%%%%%%%%%%%
\vfill\eject
\bibliography{ReferencesJP}

\begin{thebibliography}{15}
\expandafter\ifx\csname natexlab\endcsname\relax\def\natexlab#1{#1}\fi
\expandafter\ifx\csname bibnamefont\endcsname\relax
  \def\bibnamefont#1{#1}\fi
\expandafter\ifx\csname bibfnamefont\endcsname\relax
  \def\bibfnamefont#1{#1}\fi
\expandafter\ifx\csname citenamefont\endcsname\relax
  \def\citenamefont#1{#1}\fi
\expandafter\ifx\csname url\endcsname\relax
  \def\url#1{\texttt{#1}}\fi
\expandafter\ifx\csname urlprefix\endcsname\relax\def\urlprefix{URL }\fi
\providecommand{\bibinfo}[2]{#2}
\providecommand{\eprint}[2][]{\url{#2}}

\bibitem[{\citenamefont{Jamei et~al.}(2005)\citenamefont{Jamei, Kivelson, and
  Spivak}}]{Jamei:2005}
\bibinfo{author}{\bibfnamefont{R.}~\bibnamefont{Jamei}},
  \bibinfo{author}{\bibfnamefont{S.}~\bibnamefont{Kivelson}}, \bibnamefont{and}
  \bibinfo{author}{\bibfnamefont{B.}~\bibnamefont{Spivak}},
  \bibinfo{journal}{Physical Review Letters} \textbf{\bibinfo{volume}{94}},
  \bibinfo{eid}{056805} (\bibinfo{year}{2005}).

\bibitem[{\citenamefont{Horowitz
  et~al.}(2004{\natexlab{a}})\citenamefont{Horowitz, Perez-Garcia, and
  Piekarewicz}}]{Horowitz:2004yf}
\bibinfo{author}{\bibfnamefont{C.~J.} \bibnamefont{Horowitz}},
  \bibinfo{author}{\bibfnamefont{M.~A.} \bibnamefont{Perez-Garcia}},
  \bibnamefont{and}
  \bibinfo{author}{\bibfnamefont{J.}~\bibnamefont{Piekarewicz}},
  \bibinfo{journal}{Phys. Rev.} \textbf{\bibinfo{volume}{C69}},
  \bibinfo{pages}{045804} (\bibinfo{year}{2004}{\natexlab{a}}),
  \eprint{astro-ph/0401079}.

\bibitem[{\citenamefont{Horowitz
  et~al.}(2004{\natexlab{b}})\citenamefont{Horowitz, Perez-Garcia, Carriere,
  Berry, and Piekarewicz}}]{Horowitz:2004pv}
\bibinfo{author}{\bibfnamefont{C.~J.} \bibnamefont{Horowitz}},
  \bibinfo{author}{\bibfnamefont{M.~A.} \bibnamefont{Perez-Garcia}},
  \bibinfo{author}{\bibfnamefont{J.}~\bibnamefont{Carriere}},
  \bibinfo{author}{\bibfnamefont{D.~K.} \bibnamefont{Berry}}, \bibnamefont{and}
  \bibinfo{author}{\bibfnamefont{J.}~\bibnamefont{Piekarewicz}},
  \bibinfo{journal}{Phys. Rev.} \textbf{\bibinfo{volume}{C70}},
  \bibinfo{pages}{065806} (\bibinfo{year}{2004}{\natexlab{b}}),
  \eprint{astro-ph/0409296}.

\bibitem[{\citenamefont{Horowitz et~al.}(2005)\citenamefont{Horowitz,
  Perez-Garcia, Berry, and Piekarewicz}}]{Horowitz:2005zb}
\bibinfo{author}{\bibfnamefont{C.~J.} \bibnamefont{Horowitz}},
  \bibinfo{author}{\bibfnamefont{M.~A.} \bibnamefont{Perez-Garcia}},
  \bibinfo{author}{\bibfnamefont{D.~K.} \bibnamefont{Berry}}, \bibnamefont{and}
  \bibinfo{author}{\bibfnamefont{J.}~\bibnamefont{Piekarewicz}},
  \bibinfo{journal}{Phys. Rev.} \textbf{\bibinfo{volume}{C72}},
  \bibinfo{pages}{035801} (\bibinfo{year}{2005}), \eprint{nucl-th/0508044}.

\bibitem[{\citenamefont{Grotendorst et~al.}(2002)\citenamefont{Grotendorst,
  Marx, and Muramatsu}}]{Grotendorst:2002}
\bibinfo{editor}{\bibfnamefont{J.}~\bibnamefont{Grotendorst}},
  \bibinfo{editor}{\bibfnamefont{D.}~\bibnamefont{Marx}}, \bibnamefont{and}
  \bibinfo{editor}{\bibfnamefont{A.}~\bibnamefont{Muramatsu}}, eds.,
  \emph{\bibinfo{title}{Quantum Simulations of Complex Many-Body Systems: From
  Theory to Algorithms}}, Lecture Notes (\bibinfo{publisher}{John von
  Neumann-Institut für Computing (NIC)}, \bibinfo{address}{The Netherlands},
  \bibinfo{year}{2002}).

\bibitem[{\citenamefont{Wilets et~al.}(1977)\citenamefont{Wilets, Henley,
  Kraft, and MacKellar}}]{Wilets:1977}
\bibinfo{author}{\bibfnamefont{L.}~\bibnamefont{Wilets}},
  \bibinfo{author}{\bibfnamefont{E.~M.} \bibnamefont{Henley}},
  \bibinfo{author}{\bibfnamefont{M.}~\bibnamefont{Kraft}}, \bibnamefont{and}
  \bibinfo{author}{\bibfnamefont{A.~D.} \bibnamefont{MacKellar}},
  \bibinfo{journal}{Nucl. Phys.} \textbf{\bibinfo{volume}{A282}},
  \bibinfo{pages}{341} (\bibinfo{year}{1977}).

\bibitem[{\citenamefont{Dorso et~al.}(1987)\citenamefont{Dorso, Duarte, and
  J.Randrup}}]{Dorso:1987}
\bibinfo{author}{\bibfnamefont{C.}~\bibnamefont{Dorso}},
  \bibinfo{author}{\bibfnamefont{S.}~\bibnamefont{Duarte}}, \bibnamefont{and}
  \bibinfo{author}{\bibnamefont{J.Randrup}}, \bibinfo{journal}{Phys. Lett. B}
  \textbf{\bibinfo{volume}{188}}, \bibinfo{pages}{287} (\bibinfo{year}{1987}).

\bibitem[{\citenamefont{Boal and Glosli}(1988)}]{Boal:1988}
\bibinfo{author}{\bibfnamefont{D.~H.} \bibnamefont{Boal}} \bibnamefont{and}
  \bibinfo{author}{\bibfnamefont{J.~N.} \bibnamefont{Glosli}},
  \bibinfo{journal}{Phys. Rev. C} \textbf{\bibinfo{volume}{38}},
  \bibinfo{pages}{1870} (\bibinfo{year}{1988}).

\bibitem[{\citenamefont{Peilert et~al.}(1992)\citenamefont{Peilert, Konopka,
  St\"ocker, Greiner, Blann, and Mustafa}}]{Peilert:1992}
\bibinfo{author}{\bibfnamefont{G.}~\bibnamefont{Peilert}},
  \bibinfo{author}{\bibfnamefont{J.}~\bibnamefont{Konopka}},
  \bibinfo{author}{\bibfnamefont{H.}~\bibnamefont{St\"ocker}},
  \bibinfo{author}{\bibfnamefont{W.}~\bibnamefont{Greiner}},
  \bibinfo{author}{\bibfnamefont{M.}~\bibnamefont{Blann}}, \bibnamefont{and}
  \bibinfo{author}{\bibfnamefont{M.~G.} \bibnamefont{Mustafa}},
  \bibinfo{journal}{Phys. Rev. C} \textbf{\bibinfo{volume}{46}},
  \bibinfo{pages}{1457} (\bibinfo{year}{1992}).

\bibitem[{\citenamefont{Maruyama et~al.}(1998)}]{Maruyama:1997rp}
\bibinfo{author}{\bibfnamefont{T.}~\bibnamefont{Maruyama}}
  \bibnamefont{et~al.}, \bibinfo{journal}{Phys. Rev.}
  \textbf{\bibinfo{volume}{C57}}, \bibinfo{pages}{655} (\bibinfo{year}{1998}),
  \eprint{nucl-th/9705039}.

\bibitem[{\citenamefont{Watanabe et~al.}(2003)\citenamefont{Watanabe, Sato,
  Yasuoka, and Ebisuzaki}}]{Watanabe:2003}
\bibinfo{author}{\bibfnamefont{G.}~\bibnamefont{Watanabe}},
  \bibinfo{author}{\bibfnamefont{K.}~\bibnamefont{Sato}},
  \bibinfo{author}{\bibfnamefont{K.}~\bibnamefont{Yasuoka}}, \bibnamefont{and}
  \bibinfo{author}{\bibfnamefont{T.}~\bibnamefont{Ebisuzaki}},
  \bibinfo{journal}{Phys. Rev. C} \textbf{\bibinfo{volume}{68}},
  \bibinfo{pages}{035806} (\bibinfo{year}{2003}).

\bibitem[{\citenamefont{Pathria}(1996)}]{Pathria:1996}
\bibinfo{author}{\bibfnamefont{R.~K.} \bibnamefont{Pathria}},
  \emph{\bibinfo{title}{Statistical Mechanics}}
  (\bibinfo{publisher}{Butterworth-Heinemann}, \bibinfo{address}{Oxford},
  \bibinfo{year}{1996}), \bibinfo{edition}{2nd} ed.

\bibitem[{\citenamefont{Fetter and Walecka}(1971)}]{Fetter:1971}
\bibinfo{author}{\bibfnamefont{A.~L.} \bibnamefont{Fetter}} \bibnamefont{and}
  \bibinfo{author}{\bibfnamefont{J.~D.} \bibnamefont{Walecka}},
  \emph{\bibinfo{title}{Quantum Theory of Many Particle Systems}}
  (\bibinfo{publisher}{McGraw-Hill, New York}, \bibinfo{year}{1971}).

\bibitem[{\citenamefont{Vesely}(2001)}]{Vesely:2001}
\bibinfo{author}{\bibfnamefont{F.~J.} \bibnamefont{Vesely}},
  \emph{\bibinfo{title}{Computational Physics: An Introduction}}
  (\bibinfo{publisher}{Kluwer Academic}, \bibinfo{address}{New York},
  \bibinfo{year}{2001}).

\bibitem[{\citenamefont{Neumann and Fai}(1994)}]{Neumann:1994bn}
\bibinfo{author}{\bibfnamefont{J.~J.} \bibnamefont{Neumann}} \bibnamefont{and}
  \bibinfo{author}{\bibfnamefont{G.~I.} \bibnamefont{Fai}},
  \bibinfo{journal}{Phys. Lett.} \textbf{\bibinfo{volume}{B329}},
  \bibinfo{pages}{419} (\bibinfo{year}{1994}), \eprint{nucl-th/9404025}.

\end{thebibliography}
\end{document}